\documentstyle[12pt]{article}
\newskip\humongous \humongous=0pt plus 1000pt minus 1000pt
\def\caja{\mathsurround=0pt}

\def\eqalign#1{\,\vcenter{\openup1\jot \caja
        \ialign{\strut \hfil$\displaystyle{##}$&$
        \displaystyle{{}##}$\hfil\crcr#1\crcr}}\,}
\newif\ifdtup

\def\be{\begin{equation}}
\def\ee{\end{equation}}
\def\ba{\begin{eqnarray}}
\def\ea{\end{eqnarray}}

\begin{document}
\renewcommand{\theequation}{\thesection.\arabic{equation}}
\newcommand{\beq}{\begin{equation}}
\newcommand{\eeq}[1]{\label{#1}\end{equation}}
\newcommand{\ber}{\begin{eqnarray}}
\newcommand{\eer}[1]{\label{#1}\end{eqnarray}}
\begin{titlepage}
\begin{center}

\hfill CPTH-S450.0496 \\
\hfill hep-th/9605028\\

\vskip .5in

{\large \bf Threshold effects  in open-string theory}
\vskip 1cm
{ C. Bachas \footnote{bachas@orphee.polytechnique.fr}
 and  C. Fabre  \footnote{fabre@orphee.polytechnique.fr} 
} 
\vskip .1in
{ Centre de Physique Th\'eorique, Ecole Polytechnique,
 91128 Palaiseau, FRANCE \\
}

\vskip .15in

\end{center}

\vskip .4in

\begin{center} {\bf ABSTRACT }
\end{center}

\vskip .15in
We analyze  the one-loop effective gauge-field  action in
 $Z_2$-orbifold compactifications of type-I theory.
We show how,  for non-abelian group factors, the threshold effects
are ultraviolet finite though  given entirely
by a six-dimensional field theory expression.

\begin{quotation}\noindent
\end{quotation}
\vskip1.0cm
CPTH-S450.0496\\
May 1996\\
\end{titlepage}
\vfill
\eject
\def\baselinestretch{1.2}
\baselineskip 16 pt
\noindent
 \setcounter{equation}{0}

{\bf 1. Introduction}
\vskip 0.2cm

 Threshold effects  in heterotic string theory
 \cite{Kap,DKL,KK}  have been studied
intensively in the past,
 both in relation to
 gauge coupling unification \cite{dien}, 
and  in the context of 4d  N=2 heterotic-type II duality
 \cite{vaf,etc,greg}.
In open string theory, on the other hand, they only now start
to receive  attention \cite{DL}. This is in part due to the fact
that exact
 threshold  calculations are usually done in orbifold limits, and
that the rules of orbifold compactifications for type-I theory
have proven much harder to elucidate \cite{orb,PS,GP,GJ}. 
In this paper  we will analyze the full  one-loop
 Lagrangian for slowly-varying gauge-field strengths 
 in the
$Z_2$-orbifold  models constructed recently by  Gimon
 and Polchinski
 \cite{GP}. 
 These models have N=1  supersymmetry in six dimensions, 
and a maximal gauge group $G= U(16)\times {\tilde U(16)}$,
 which can be broken by both discrete and continuous
(antisymmetric) moduli. Upon toroidal compactification to four
dimensions, one finds  N=2 supersymmetries  and extra (adjoint)
Wilson-line moduli.
 In calculating the effective gauge-field action
 we will follow
 reference  \cite{BP}, where a  similar calculation
was carried out  for toroidal compactifications of the
type-I  SO(32) superstring. Our conclusions  can be summarized as
follows:
\vskip 0.15cm

\indent   {\it (a)} the structure of ultraviolet divergences
 is identical
   in the toroidal and \hfil\break\indent\ \ \ \
 orbifold models,
   and can be traced  to tadpoles of the dilaton and 
\hfil\break\indent\ \ \ \ graviton,
   provided the background field has
   no component along one \hfil\break\indent\ \ \ \
  of the  simple  U(1) factors in six-dimensions; 
\hfil\break \indent
   {\it (b)} gauge coupling renormalization is  entirely
due  to  six-dimensional \hfil\break\indent\ \ \ \
 field-theory states, but is  ultraviolet
finite  due to the (unconventional) \hfil\break\indent\ \ \ \
 cutoff prescription dictated
by the string.

\vskip 0.15cm

To understand the above results heuristically, note  that
 the divergent parts of
 one-loop string amplitudes are 
proportional to the {\it square}  of closed-string 
 tadpoles, which 
 in a consistent theory 
 are in turn  $o(F_{\mu\nu}^2)$ each.
 The only possible  exception
to this rule comes from Green-Schwarz couplings of twisted
closed-string states,
required to cancel the anomalies  of simple
U(1) gauge-group factors \cite{Mich}, and giving rise to tadpoles
of $o(F_{\mu\nu})$.
For traceless background fields the quadratic part of the induced
Lagrangian is thus ultraviolet finite. As for the fact that
only massless six-dimensional states contribute, it follows from
the observation  that these are the only open-string
 BPS states: all other
states have the spin content of N=4 multiplets and do not therefore
renormalize the gauge coupling \cite{greg}
\footnote{We thank E. Kiritsis
for a clarification of this point.}. This last conclusion has been
also reached in a somewhat different
 context  by Douglas and Li  \cite{DL},
though  in their D-brane realization of  N=2 super
 Yang-Mills ultraviolet
finiteness of the thresholds is not explicit.

   The structure of this paper is as follows: in section 2 we will
review the one-loop calculation of the  gauge action for
toroidal compactifications of the SO(32) type-I superstring.
 In section 3 we will analyze  ultraviolet divergences, and rederive
in particular the relation between the gauge and gravitational
 couplings and  the string
scale, obtained  previously by Abe and Sakai \cite{AS,Thorn}.
The results of these two  sections are standard \cite{CC,Ark},
 but we include them as a warm up for the orbifold
calculation  which follows  in section 4.
 In section 5 we will consider  in
particular the quadratic piece of the induced action,
 and show that it  can be obtained
from 6d field theory with a subtle  cutoff prescription.
We will conclude with some comments  and perspective.

\vskip 0.6cm

{\bf 2.  Annulus, M\"obius strip and Klein bottle}
\vskip 0.2cm

We will first consider  toroidal compactifications of
the 
$ SO(32)$ type-I  theory down to four space-time dimensions,
that is on $R^4\times T^6$.
 Many technical steps are best 
illustrated in this simple context, even though maximal
unbroken  supersymmetry implies that  the $\beta$-functions
and  threshold corrections vanish. 
Besides the torus (${\cal T}$),  one-loop vacuum diagrams include
 the annulus (${\cal A}$), the
M\"obius strip (${\cal M}$) and the Klein bottle (${\cal K}$). 
These are given respectively by
$$
\eqalign{
&{\cal A}(i,j)= -{1\over 2} \int_0^\infty {dt\over t}\
{\rm Str}_{\ \atop {\rm  open \atop (i,j)}}
\  e^{-{\pi t\over 2}(k^\mu k_\mu + M^2)} \cr
  &{\cal M}(i)= -{1\over 2} \int_0^\infty {dt\over t}\
{\rm Str}_{\ \atop {\rm  open \atop (i,i)}}
\ \Omega  e^{-{\pi t\over 2}(k^\mu k_\mu + M^2)} \cr
&\ \ {\cal K}= -{1\over 2} \int_0^\infty {dt\over t}\
{\rm Str}_{\ \atop {\rm  closed}}
\ \Omega  e^{-{\pi t\over 2}(k^\mu k_\mu + M^2)} \cr
} \eqno(2.1)
$$
where 
$\Omega$ is the world-sheet reflection operator, 
$M^2$ the  mass-squared operator in four dimensions, 
$(i,j)$ are Chan-Paton labels of the open-string endpoints, and
the supertrace includes a sum over 
energy-momentum,
$$
Str = V^{(4)} \int {d^4k\over (2\pi)^4}\  
\Bigl( \sum_{bos}-\sum_{ferm}
\Bigr) \eqno(2.2)
$$
with $V^{(4)}$ the regulated volume of $R^4$.
Only open strings with identical endpoint charges contribute
to the M\"obius trace, as follows also from the fact that
the corresponding surface has a single boundary. Likewise,
only closed strings with identical left- and right-moving
excitations contribute to the Klein-bottle trace.

The expressions of these amplitudes for zero
 gauge-field strength 
and arbitrary Wilson-line backgrounds  are
$$
\eqalign{\ \ \ \ \ 
&{\cal A}(i,j)=
-{i\over 2} 
 V^{(4)} \int_0^\infty {dt\over t} (2\pi^2 t)^{-2}
\Bigl(  \sum_{a_i+a_j+^*\Gamma_6} e^{-{\pi t} p^2/2} \Bigr)  \
Z(it/2)
\cr
&{\cal M}(i)=
 {i\over 2} 
 V^{(4)} \int_0^\infty {dt\over t} (2\pi^2 t)^{-2}
\Bigl(  \sum_{2a_i+^*\Gamma_6} e^{-{\pi t} p^2/2} \Bigr) \  
Z({it/2}+{1/ 2}) \cr
&\ \ \ \ \ {\cal K}=
-{i\over 2} 
 V^{(4)} \int_0^\infty {dt\over t} (2\pi^2 t)^{-2}
 \Bigl( \sum_{^*\Gamma_6} e^{-{\pi t} p^2/2} \Bigr) \  Z(2it)
\cr} \eqno(2.3)
$$
where
$$ Z(\tau)=
{1\over  \eta^{12} ( {\tau})} \times  
\ \sum_{\alpha=2,3,4}  \  {1\over 2} s_\alpha
\theta_\alpha^4 (0\vert \tau) \eqno(2.4)
$$
is the usual open-string oscillator sum, 
 $s_3=-s_2=-s_4=1$ are the  GSO projection signs,
and we have set 
$$2\alpha^\prime = 1 \ .
$$ 
The sums inside the large parentheses run over
the internal   momentum  lattice
$\ ^*\Gamma_6$,  shifted from the origin by the 
 Wilson-line backgrounds.
Explicitly, $a_i^I$ is the eigenvalue on Chan-Patton
 state $\vert i>$
 of the constant gauge-field
background $A^I$, pointing in the $I$th direction  on the torus,
and lying in the Cartan subalgebra of SO(32). 
 Besides these shifts, the three expressions in
(2.3)
 differ only
   in the argument of the
modular functions \cite{PS}:
  the 
argument in the  Klein bottle is four times that
in the  annulus, because  closed-string Regge trajectories
 have
a  mass spacing twice as large as the corresponding
 open-string ones.
The extra  $+ {1\over 2}$ in the
argument of  the M\"obius amplitude
takes into account properly the eigenvalues of oscillator
excitations under the reflection operator $\Omega$.
The sign of ${\cal M}$ has been  chosen so that the
eigenvalue of massless 
open-string states under $\Omega$ be minus one.
Finally notice that the Klein bottle symmetrizes the Neveu-Schwarz
Neveu-Schwarz states, and antisymmetrizes the Ramond-Ramond states,
as required by space-time supersymmetry.

  The presence of a constant electromagnetic background modifies
these amplitudes in a simple way \cite{BP,Abouel}.
 We will choose  for definiteness
a (positive)  magnetic field in the $X^1$ direction 
$$ F_{23} = {\cal B}  Q \ , 
 \eqno(2.5) $$
with $Q$ a Cartan-subalgebra 
 generator normalized so that $tr_{\rm fund} Q^2 
={1\over 2}$. 
The net effect \cite{Abouel}
 of the field on the open-string spectrum is 
a shift of the oscillator frequencies of the complex
coordinate $X_2+ iX_3$ by an amount $ \epsilon$, where
$$
\pi \epsilon = {\rm arctan}( \pi q_i {\cal B}) + 
{\rm arctan}( \pi q_j  {\cal B} ) \  \eqno(2.6) 
$$
and $q_i$, $q_j$ are defined as above.
For simplicity of notation, the dependence of $\epsilon$ on the
endpoint states $i,j$ will be implicit in the sequel.
The annulus and M\"obius-strip  amplitudes  
are now given by eqs. (2.1) with the replacements
$$\eqalign{& \ \cr
&k_\mu k^\mu  \rightarrow -(k_0)^2 + (k_1)^2 +
(2n+1)\epsilon
+2\epsilon\Sigma_{23} \cr
&Str \rightarrow V^{(4)} {(q_i+q_j){\cal B}\over 2\pi}
\int {d^2k\over (2\pi)^2}\  \sum_n \
\Bigl( \sum_{bos}-\sum_{ferm}
\Bigr)\ \cr}
  \eqno(2.7)
$$
Here
 $\Sigma_{23}$ is the spin operator in the $(23)$ direction, and 
$n=0,1,..$ labels the  Landau levels whose degeneracy 
per unit area is
${(q_i+q_j){\cal B}\over 2\pi}$.  
The
above  formula \cite{FPT}
 encodes in particular the fact that all open-string
states have the same gyromagnetic ratio $g=2$. 
The  reflection operator 
$\Omega$ acts  as in the zero-field limit. 
Performing the supertraces explicitly
 leads to
the following 
expressions for the  amplitudes
$$
\eqalign{\ \ \ 
{\cal A}(i,j\vert {\cal B})= 
-{i\over 2}&  V^{(4)} \int_0^\infty {dt\over t} (2\pi^2 t)^{-2}
\ \Bigl(  \sum_{a_i+a_j+^*\Gamma_6} e^{-{\pi t} p^2/2} \Bigr)
{1\over  \eta^{12} ( {it\over 2})} \times \cr \times 
{ i \over 2}&{ (q_i+q_j)}{\cal B}  t \  
{\theta_1^\prime (0\vert {it\over 2})
\over \theta_1 ({i\epsilon t\over 2} \vert {it\over 2})} 
\ \sum_{\alpha=2,3,4}  \  {1\over 2} s_\alpha
  \theta_\alpha ({i\epsilon t\over 2}\vert {it\over 2})
\theta_\alpha^3 (0\vert {it\over 2})
 \cr}\eqno(2.8)
$$
and 
$$\eqalign{\ \  
{\cal M}(i\vert {\cal B})= 
{i\over 2}&  V^{(4)} \int_0^\infty {dt\over t} (2\pi^2 t)^{-2}
\ \Bigl(  \sum_{2a_i+^*\Gamma_6} e^{-{\pi t} p^2/2} \Bigr)
{1\over  \eta^{12} ( {it\over 2}+{1\over 2})} \times \cr \times 
i{\cal B} q_i t \ & 
{\theta_1^\prime (0\vert {it\over 2}+{1\over 2})
\over \theta_1 ({i\epsilon t\over 2} \vert {it\over 2}+{1\over 2})} 
\ \sum_{\alpha=2,3,4}  \  {1\over 2} s_\alpha
  \theta_\alpha ({i\epsilon t\over 2}\vert {it\over 2}+{1\over 2})
\theta_\alpha^3 (0\vert {it\over 2}+{1\over 2}) \ 
 \cr}\eqno(2.9)
$$
The Klein bottle is of course unmodified.
The reader can verify as a check  that these amplitudes reduce 
to  (2.4)  in the ${\cal B} = 0$  limit.

\vskip 0.1cm

The
 full one-loop free energy reads
$$
i{\cal F}  = {1\over 2}({\cal T} + {\cal K})
 + {1\over 2} \sum_{ij} {\cal A}(i,j\vert {\cal B})
+ {1\over 2} \sum_i {\cal M}(i\vert {\cal B}) \ . \eqno(2.10)
$$
with ${\cal T}$ the contribution of the torus.
The
 $o({\cal B}^2)$ Maxwell term 
 vanishes for toroidal
 compactifications
 as a result of $N=4$
supersymmetry. In the form (2.8)-(2.9) of the amplitudes, this
 follows from
 the well known identity
$$\sum_{\alpha} s_\alpha \theta^2_\alpha(\nu\vert \tau)
\theta^2_\alpha(0\vert \tau)
=0 \  \eqno(2.11)
$$
and the fact that $\theta_2$, $\theta_3$ and $\theta_4$ are
even functions of their (first) argument. 
Alternatively, from eqs. (2.1)-(2.7) one obtains the following
contribution  of a supermultiplet to ${\cal F}$:
$$
\eqalign{
\ V^{(4)} \ \int_0^\infty {dt\over t}\ e^{-{\pi t\over 2} M^2}\  
\Biggl\{ -{1\over 8\pi^4t^2}\ &  str ({\bf 1})\  +\  
{{\cal B}^2 (q_i+q_j)^2   \over 16\pi^2} 
\ str ( {{\bf 1}\over 12} -\Sigma_{23}^2) \cr &
-
{{\cal B}^2 (q_i^2+q_j^2 -q_i q_j)\over 24\pi^2 t^2}\  str({\bf 1})
\  + o({\cal B}^4)
\Biggr\} \cr}  \eqno(2.12)
$$
where $str$ sums over spin states of 
the  supermultiplet. 
The first two terms in the above expansion are the
one-loop  field-theory corrections to the vacuum energy and 
Maxwell action, while the third term is due to  the stringy nature
of the particles. 
All of  these terms vanish identically
for complete  $N=4$ supermultiplets, consistently with the fact that
such multiplets do not renormalize the gauge coupling.

\vskip 1cm
{\bf 3. Ultraviolet divergences}
\vskip 0.2cm

In heterotic string theory ultraviolet finiteness
at one loop follows from
the restriction of the integration  over all  world-sheet  tori
to a single  fundamental
domain. This presupposes conformal invariance, 
 or equivalently the absence of
classical tadpoles. 
Since a  background $F_{\mu\nu}$ does  give  rise to  tadpoles for
the graviton and dilaton, a complete  background-field calculation
requires an 
appropriately curved space-time \cite{KK}.
This is also true in  type-I theory, but massless
 closed-string tadpoles
may now manifest themselves as ultraviolet divergences from
small holes, rather than as a  violation of conformal invariance
in the world-sheet interior. Such divergences should be cancelled
by a generalized Fishler-Susskind mechanism \cite{FS,CC,Ark}. As we
will see this only affects the results at $o({\cal B}^4)$, so that
the threshold calculation can be consistently performed in flat
space-time.

In order to study the ultraviolet behaviour of the amplitudes
we must rewrite them in terms of the 
 proper time  in the (transverse)
closed string channel \cite{l}.
 We normalize this proper
time through   the closed-string
propagator, 
$$
\Delta_{closed} = {\pi\over 2}
 \int_0^\infty dl\  e^{-{\pi l\over 2}(k^\mu k_\mu +
M^2_{closed})} \  \eqno(3.1)
$$
for scalar states.
The relation between $l$ and the proper time in the direct channel
 is  different
 for each  surface \footnote{Our conventions differ from those
of Gimon  and Polchinski \cite{GP}.
 With our normalization of the direct-channel
proper time,  a cutoff
 $t> \Lambda^{-2}$  is equivalent to a universal
 momentum cutoff on all
open and closed-string states.} 
$$
 l = \cases{& $1/t$ \ \ annulus ;\cr
&$1/4t$ \ \ M\"obius ;\cr
& $1/4t$ \ \ Klein bottle .\cr} \eqno(3.2)
$$
To study the $l\rightarrow\infty$ limit  we must
use the modular properties of the elliptic functions. 
For the annulus we use their transformation under
$\tau = it/2 \rightarrow -1/\tau=2il$, together with the
Poisson resummation formula
 to find
$$\eqalign{ 
{\cal A}(i,j\vert {\cal B})= 
-{i\over 32} &  v^{(4)} v^{(6)}  {(q_i+q_j){\cal B}}
\ \int_0^\infty  dl \ \Biggl(
 \sum_{w\in\Gamma_6} e^{-w^2 l /2\pi -i (a_i+a_j)\cdot w } \Biggr)
 \times \cr & \times
  \eta^{-12} ({2il} )\ 
{\theta_1^\prime (0\vert{2il})
\over \theta_1 ({\epsilon} \vert {2il})} 
\ \sum_{\alpha=2,3,4}  \ {1\over 2}  s_\alpha
  \theta_\alpha ({\epsilon}\vert{2il}  )
\theta_\alpha^3 (0\vert {2il} ) \cr} 
\eqno(3.3)
$$
where   
$V^{(6)}$  is the volume of the compact six-torus,
 $\Gamma_6$ the (winding)  lattice of points
identified with the origin, and 
$$v^{(d)}= V^{(d)}/(2\pi\sqrt{\alpha^\prime})^{d} \ . $$
Likewise for the M\"obius strip we use the sequence of
modular transformations
$$
\tau = {it\over 2}+{1\over 2} \rightarrow -1/\tau
 \rightarrow -1/\tau+2
\rightarrow (1/\tau -2)^{-1}= {2il} -{1\over 2}
$$
with the result

$$\eqalign{ 
{\cal M}(i\vert  {\cal B} )&= 
2 i\  { v^{(4)} v^{(6)}} { q_i {\cal B} }
\ \int_0^\infty  dl \ 
\Biggl(
 \sum_{w\in\Gamma_6} e^{-2 w^2 l /\pi -2i a_i\cdot w } \Biggr) \
  \eta^{-12} (2il -{1\over 2})
 \times
 \cr
& \times  
{\theta_1^\prime (0\vert 2il  -{1\over 2} )
\over \theta_1 ({\epsilon\over 2} \vert 2il  -{1\over 2})} 
\ \sum_{\alpha=2,3,4}  \ {1\over 2}  s_\alpha
  \theta_\alpha ({\epsilon\over 2 }\vert 2il  -{1\over 2}  )\ 
\theta_\alpha^3 (0\vert 2il  -{1\over 2} )
 \cr }
 \eqno(3.4)
$$
Finally,  after a change of
variables  $\tau = 2it \rightarrow -1/\tau = 2il$,
the Klein bottle amplitude  takes the form
$$
{\cal K} =  
-32i\    v^{(4)}v^{(6)}  
 \int_0^\infty dl
\ \Biggl( \sum_{w\in\Gamma_6} e^{-2 w^2 l/\pi } \Biggr)
\times   \eta^{-12} (2il)   
\sum_{\alpha=2,3,4}{1\over 2} s_\alpha
\theta_\alpha^4 (0\vert 2il)  \ . \eqno(3.5)
$$
\vskip 0.2cm

 Using the appropriate expansions
of the elliptic functions and some simple trigonometry, 
 we can extract the infrared divergences
of these expressions   with the result

$$\eqalign{
{\cal A}(i,j\vert {\cal B}) \sim 
{i\over 4} v^{(4)} v^{(6)}  
& \int^\infty dl\ \Biggl[\   
\underbrace{1- q_iq_j \pi^2 {\cal B}^2}_{RR}\ + 
 \cr
 +&
\underbrace{
 {{1\over 2}  (q_i+q_j)^2\pi^2 {\cal B}^2 \over  
\sqrt{1+q_i^2 \pi^2 {\cal B}^2} \sqrt{1+q_j^2\pi^2 {\cal B}^2}}
-  \sqrt{1+q_i^2 \pi^2 {\cal B}^2}
 \sqrt{1+q_j^2 \pi^2 {\cal B}^2}
}_{NS-NS}\ 
\Biggr] \cr
\ & \cr
{\cal M}(i\vert {\cal B}) \sim 
-16i\  v^{(4)} v^{(6)}   
&  \int^\infty dl \times\ 
\Biggl[ \  
\underbrace{1}_{RR}\  +\  
\underbrace{
{1\over 2}  { q_i^2\pi^2 {\cal B}^2 \over  
\sqrt{1+q_i^2\pi^2 {\cal B}^2}}
-  \sqrt{1+q_i^2 \pi^2 {\cal B}^2}
}_{NS-NS}\ 
\Biggr] 
\cr
\ & \cr
{\cal K}({\cal B}) \sim 
256i\   v^{(4)} v^{(6)}  
&  \int^\infty dl \times\ 
\Biggl[\  
\underbrace{
1
}_{R-R}\ \ 
-\ \  
\underbrace{1}_{NS-NS}
\Biggr] \cr} 
 \eqno(3.6)
$$
We have identified in the above expressions the contributions from
the Ramond-Ramond or  Neveu-Schwarz   
intermediate closed-string states. These
correspond to the $\alpha=2$
and $\alpha=3,4$  terms, respectively,  of the spin structure sums 
in the $l$-channel. 

 The RR piece of the ultraviolet divergence is polynomial in
the background field. This is because  it comes from tadpoles
in  the Wess-Zumino part of the effective action \cite{Li,Mich}
which is the integral of 10-forms made out of $F_{\mu\nu}$ and
the  antisymmetric massless RR  tensors. The constant part
is  a tadpole
for the unphysical RR 10-form \cite{CC}, 
which  cancels in the
full free energy, eq. (2.10), after summing over the 32 Chan-Patton
charges. The quadratic term,  proportional to $q_i q_j$, also
cancels after summing over opposite charges or, in modern language,
 over
conjugate pairs of 9-branes. This is consistent with the fact that
the potential coupling 
$\int (tr F) \wedge A^{(8)}$ is here absent, both because 
 the (high-energy)  gauge group
contains no  simple $U(1)$ factors, and also because the
 RR  
8-form is projected out of the spectrum  by the
 orientation reversal.

 Consider next  the   Neveu-Schwarz piece of the
 ultraviolet divergence. 
This comes  from  tadpoles of the graviton, dilaton
 and 2-index antisymmetric
 tensor, which
 couple to  the background field through a  Dirac-Born-Infeld
action \cite{BI,CC}.
 The constant term
vanishes  for SO(32) gauge group, consistently with the
fact  that the
vacuum energy in the absence of a magnetic
 field should be  zero. The
2-index  tensor
$B_{\mu\nu}$ is   projected out of the spectrum by 
orientation reversal,
 consistently again  with the fact that the
piece  proportional to  $q_i q_j$ cancels after summing over
opposite endpoint charges. 
 The first physical divergence appears thus at
order $o({\cal B}^4)$, and comes from an  on-shell
graviton  or a  dilaton. As a check of our formulae  let us
compute this directly from the
  effective open-string  action
$$
S = \int d^{10}x \sqrt{-g} \Biggl[ 
-{1\over 2\kappa_{(10)}^2}  R + {1\over 16\kappa_{(10)}^2}
 (\partial_\mu\phi)^2  + {1\over 2g_{(10)}^2}  e^{\phi/4}
 tr F_{\mu\nu} F^{\mu\nu} \Biggr]
\eqno(3.7)
$$
where $\kappa_{(10)}$ and $g_{(10)}$ are the gravitational and
gauge couplings, and the $SO(32)$ generators are normalized
to $tr(t^a t^b) = {1\over 2} \delta^{ab}$.
 A constant magnetic field
in flat space with  $<\phi>=0$, 
gives rise to  an energy-momentum tensor
$$ j_{\mu\nu} = - {1\over g_{(10)}^2} tr
\Bigl( F_{\mu\rho} F^{\ \rho}_{\nu} 
-  {1\over 4} \eta_{\mu\nu}  F_{\rho\sigma} F^{\rho\sigma}
 \Bigr)\ ,  \eqno(3.8)
$$
as well as to  a source for the dilaton
$$
j_{\phi}=  {1\over 8 g_{(10)}^2}
 tr\ F_{\mu\nu}F^{\mu\nu} \ . \eqno(3.9)
$$
Using the graviton propagator in the De Donder
gauge \cite{Velt},
$$
{1\over 2\kappa_{(10)}^2} \Delta^{\mu\nu,\rho\sigma} =
(\eta^{\mu\rho}\eta^{\nu\sigma}+ \eta^{\mu\sigma}\eta^{\nu\rho}
-{1\over 4} \eta^{\mu\nu} \eta^{\rho\sigma}) {i\over k^2}
\ , \eqno(3.10)
$$
one derives easily the following
 infrared contribution to vacuum energy
due to the above  tadpoles,
$$\eqalign{ 
\ \ \ {\cal F}  \sim -\kappa_{(10)}^2 V^{(10)}  \Bigl[
4 j_{\phi}^2 + 2 j_{\mu\nu}j^{\mu\nu} - & {1\over 4} 
(j_\mu^{\ \mu})^2 \Bigr] \times {\pi\over 2}  \int^\infty dl \cr
& \sim - {3\over 4} {\kappa_{(10)}^2 \over g_{(10)}^4}
V^{(10)} {\cal B}^4  \times {\pi\over 2}
  \int^\infty dl  \cr} \eqno(3.11)
$$
To compare with the result of the string calculation, we must
expand the annulus and M\"obius 
to order ${\cal B}^4$, and perform the summation over endpoint
charges. For a generic normalized
generator   there are
two charged  endpoint states, $q_1=-q_2 = {1\over 2}$, while 
$q_3= ...  = q_{32}=0$. After some
 straightforward algebra one  finds  agreement with eq. (3.11)
  provided
$$ g_{(10)}^4 = 2^9 \pi^7 \kappa_{(10)}^2 
(2\alpha^\prime)^2 \eqno(3.12)
$$
where we have here restored correct mass  units 
\footnote{More appropriately we should have
 replaced $1/\alpha^\prime$
by the mass squared of the 
 first open-string
excitation. Equation (3.12) would then be manifestly
 invariant under Weyl rescalings of the metric.} .
This relation between the gravitational and gauge
 coupling constants
has been  derived previously by  Abe and Sakai \cite{AS}, and 
in the case of  the
bosonic string by Shapiro and  Thorn, and Dai and Polchinski
 \cite{Thorn}.
In  terms of
four-dimensional couplings it reads
$$  g_{(4)}^4 \alpha^\prime = 16\pi 
\kappa_{(4)}^2
 { (4\pi^2\alpha^\prime)^3 \over V^{(6)} }
\eqno(3.13) $$
Contrary to what happens for the heterotic string \cite{Gins},
the 
compactification volume enters non-trivially here.
Thus in open string theory the string  scale is not
  irrevocably tied  to the Planck scale at tree level,
a  fact that has received some attention
  recently in refs. \cite{Witt}.

  The basic lesson to retain here is that
 the quadratic part of the
effective action is ultraviolet convergent in flat space-time,
without evoking  space-time supersymmetry. This will
continue to hold, with a slight caveat,   in the 
 orbifold compactification to which we turn now.

\vskip 0.9cm

{\bf 4.  $Z_2$ orbifold}
\vskip 0.2cm

 Let us first briefly recall the upshot of the
 analysis by Gimon  and
Polchinski \cite{GP}. The type-I theory on $R^6\times T^4/Z^2$
contains  untwisted and twisted closed strings, 
 as well as open strings of three different types:
those with freely moving endpoints (NN or 99), those whose
endpoints are stuck on some 5-branes transverse to the orbifold
(DD or 55), and  those with one endpoint stuck and one
moving freely (DN or 59). Consistency fixes both the number
of 9-branes and the number of 5-branes to be 32. 
It also fixes the action of the orientation reversal
$(\Omega )$ and orbifold-twist $({\cal R})$
on the open-string end-point states. 
This action does not mix Neumann with Dirichlet endpoints, 
and  can be described in
appropriate bases by the four $32\times 32$ matrices
$$
\gamma_{\Omega,9} = {\bf 1} \ , \ \ \ 
\gamma_{{\cal R},9} = \gamma_{{\cal R},5} =
\gamma_{{\Omega},5} =
 \left( \matrix{ 0 & i{\bf 1}\cr -i{\bf 1} & 0\cr}\right)
\eqno(4.1)
$$
The massless spectrum of this theory in six dimensions
 has\hfil\break
\indent  {\it (i)}  the N=1 supergravity, one tensor
 and four  gauge-singlet
hypermultiplets from the untwisted  closed-string
 sector,\hfil\break
\indent {\it (ii)} sixteen gauge-singlet
  hypermultiplets, one from each
fixed-point of the orbifold, in the twisted closed-string
 sector\hfil\break 
\indent {\it (iii)}  $ U(16)$  vector
 multiplets and two  hypermultiplets in the
antisymmetric $ 120$
 representation from the NN sector,\hfil\break
\indent {\it (iv)} identical  content, i.e.
 an extra ${\tilde U(16)}$
gauge group and two  antisymmetric hypermultiplets,
 from the DD sector,
and \hfil\break
\indent  {\it (v)} one hypermulitiplet transforming in the
representation  $(16,16)$ of  the full gauge group
and coming  from the DN sector.\hfil\break
\indent  Notice that each  twisted-sector
hypermultiplet  contains  a  RR 4-form field $ C^{(I)}$ 
localized at the $I$th fixed point of the orbifold, which plays
a special role in what follows.

 This model has a T-duality, which interchanges NN and DD sectors
and hence also the two U(16) gauge groups. Without losing
 generality, we may thus  restrict ourselves to background fields
${\cal B}Q$  arising
from the NN sector. The 
 antisymmetric hypermultiplets are six-dimensional moduli.
They  have a simple geometric interpretation in the DD sector
 \cite{GP}, where
they correspond to motion of a pair of 5-branes with  their mirror
image away from a fixed-point of the orbifold. Together with 
the discrete moduli,  that  correspond to jumps of a 5-brane  pair
  between fixed points,
 these  can be used
to break the gauge symmetry to various  unitary and symplectic
components. 
To simplify the calculation we will turn off all six-dimensional
moduli in what follows, i.e. we will assume
maximal unbroken gauge symmetry in six dimensions.
The more general calculation  presents
no real technical difficulty. We 
will  furthermore choose to  work on
$$ R^4 \times T^2 \times T^4/Z_2 $$
so that we may break the gauge group by Wilson-line moduli in four
dimensions.
We denote  volumes as $V^{(4)}$,
$V^{(2)}$ and $V^{orbifold}$, the latter being half the volume of
the corresponding torus.
There is one final detail to settle: in the basis of Chan-Patton
charges of ref. \cite{GP}, the wavefunctions of NN gauge bosons
take the form
$$
\lambda = \left( \matrix{ A& S \cr -S & A \cr} \right)
\eqno(4.1)
$$
where $A$ and $S$ are arbitrary $16\times 16$ antisymmetric and
symmetric matrices. Since in order to perform our calculation we
 need  
 to diagonalize  the background gauge field, we 
will make the  unitary change of basis of Neumann endpoint states
$$
U = {1\over\sqrt{2}} \left( \matrix{{\bf  1}
 &\ i{\bf  1} \cr {\bf  1}& -i{\bf  1}\cr} \right)
 \eqno(4.2)
$$
In this  new basis the $16$ and
 ${\overline { 16}}$ representations of
the gauge group are disentangled, and the orbifold-twist operator
acts as a simple sign, 
 $$s_{ij}= -1\ \ \ {\rm  or} \ \ \ +1 \ , $$
 according to whether
the end-point states
$\vert i>$ and $\vert j>$  belong to  the same or
 to conjugates representations.

\vskip 0.1cm

We are now ready to proceed with the calculation of the one-loop
free energy, which is the sum of contributions from the various
sectors,
$$
{\cal F}^{orbifold} = {\cal F}_{closed}+
{\cal F}_{NN}+{\cal F}_{ND}+{\cal F}_{DD} \eqno(4.3)
$$
Since only Neumann endpoints couple to the background field, 
we can ignore 
${\cal F}_{closed}$ and ${\cal F}_{DD}$ which  vanish
by space-time supersymmetry\footnote{Consistency of the theory
requires of course ultraviolet finiteness for every individual
cross channel. To check it one needs the explicit forms
for these amplitudes, before enforcing supersymmetry identities.
Since our conventions differ somewhat from those of
Gimon  and Polchinski, we  give these expressions 
 in the appendix for completeness.}. 
The remaining two contributions read
$$
i{\cal F}_{NN} =  
{1\over 4}
 \Biggl\{
  \sum_{ij} {\cal A}(i,j\vert {\cal B})
+\sum_{ij}  {\cal A}^{({\cal R})} (i,j\vert {\cal B}))
+ \sum_i {\cal M}(i\vert {\cal B})
+ \sum_i  {\cal M}^{({\cal R})} (i\vert {\cal B})
\Biggr\}
$$
and
$$
i{\cal F}_{ND} = {1\over 2} \times 32 \sum_i
 {\cal A}_{ND}(i\vert {\cal B})
\eqno(4.4)
$$
where the superscript ${\cal R}$ here indicates the insertion of
the orbifold-twist operator inside a trace.
Only a single ampitude contributes in the ND sector: indeed,
there are no M\"obius diagrams, 
 since the
action of $\Omega$ does not mix Neumann and Dirichlet states,
and  ${\cal A}_{ND}^{({\cal R})} \propto
 tr(\gamma_{5,{\cal R}}) = 0$,
by eq. (4.1). 
Notice also that the overall factor in front of ${\cal A}_{ND}$
 takes into account the multiplicity
of 5-branes, as well as the two possible orientations of a ND
string.

The NN annulus and M\"obius, without insertion of the ${\cal R}$
operator, are given by eqs. (2.8-9), with 
$\ ^*\Gamma_6 =\ ^*\Gamma_2\oplus\ ^*\Gamma_4$ being
 the direct sum of 
the lattices of momenta on the two-torus and the orbifold, and
with the Wilson-lines shifitng only the former  momenta.
The other two NN amplitudes read
$$
\eqalign{\ \ \ 
{\cal A}^{(R)} (i,j\vert {\cal B})= 
-{i\over 2}  V^{(4)}& s_{ij}
 \int_0^\infty {dt\over t} (2\pi^2 t)^{-2} 
 \Bigl(  \sum_{a_i+a_j+^*\Gamma_2} e^{-{\pi t} p^2/2} \Bigr)
\
{1\over  \eta^{8} }
 {4 \eta^{2} \over \theta_2^2}
\times \cr \times 
{ i \over 2} &{ (q_i+q_j)}{\cal B}  t \  
{\theta_1^\prime (0)
\over \theta_1 ({i\epsilon t\over 2})} 
{1\over 2}  \Biggl(
  \theta_3 ({i\epsilon t\over 2})
\theta_3 
\theta_4^2  - 
 \theta_4 ({i\epsilon t\over 2})
\theta_4 
\theta_3^2 
\Biggr)
 \cr}\eqno(4.5)
$$
and
$$
\eqalign{\ \ \ 
{\cal M}^{(R)}(i\vert  {\cal B})= 
- {i\over 2}  V^{(4)}
& \int_0^\infty {dt\over t} (2\pi^2 t)^{-2}\ 
 \Bigl(  \sum_{2a_i+^*\Gamma_2} e^{-{\pi t} p^2/2} \Bigr)
{1\over  \eta^{8} }
 {4 \eta^{2} \over \theta_2^2}
\times \cr \times 
{ i}&{ q_i}{\cal B}  t \  
{\theta_1^\prime (0)
\over \theta_1 ({i\epsilon t\over 2})} 
{1\over 2}  \Biggl(
  \theta_3 ({i\epsilon t\over 2})
\theta_3 
\theta_4^2  - 
 \theta_4 ({i\epsilon t\over 2})
\theta_4 
\theta_3^2 
\Biggr)
 \cr} \eqno(4.6)
$$
The modular parameter of the elliptic
 functions inside the integrals
 is 
$\tau={it\over 2}$ for the annulus
and $\tau={it\over 2}+{1\over 2}$ for the M\"obius strip,
 and
the first argument of the theta functions is by default zero.
In deriving the
 above expressions
we made use of  the fact that
$4\eta^2/\theta_2^2$ is the correctly normalized
contribution of the four bosonic orbifold coordinates
twisted in the time direction, 
 that the (open-string)  Ramond sector does not contribute
 because of fermionic zero modes,
 and finally that $s_{ii} = -1$.
Likewise the  ND amplitude reads
$$
\eqalign{
{\cal A}_{ND}(i\vert {\cal B})= 
-{i\over 2}  V^{(4)}
 &\int_0^\infty {dt\over t} (2\pi^2 t)^{-2}
 \Bigl(  \sum_{a_i+^*\Gamma_2} e^{-{\pi t} p^2/2} \Bigr)
{1\over  \eta^{8}} {\eta^2\over \theta_4^2}
  \times \cr \times 
{ i \over 2}&{ q_i}{\cal B}  t \  
{\theta_1^\prime 
\over \theta_1 ({i\epsilon t\over 2})}
{1\over 2} 
 \Biggl(
  \theta_3 ({i\epsilon t\over 2})
\theta_3 
\theta_2^2  - 
 \theta_2 ({i\epsilon t\over 2})
\theta_2 
\theta_3^2 
\Biggr) 
 \cr}\eqno(4.7)
$$
where we have taken here into account that ND
bosonic  coordinates have half-integer frequencies, while their
fermionic partners have integer or half-integer frequencies
in the Neveu-Schwarz or Ramond  sectors.

\vskip 0.2cm

Let us summarize the calculation as
$${\cal F}^{orbifold}({\cal B})  =
 {1\over 2} {\cal F}^{toroidal}({\cal B}) + 
\delta {\cal F}({\cal B}) \ , \eqno(4.8)
$$
where ${\cal F}^{toroidal }$ is the induced
 action of the theory before
the orbifold projection, and $\delta{\cal F}$ is given by sums
of  the  
 amplitudes (4.5-7) over endpoint states. What we will now show is
that $\delta{\cal F}$ has no ultraviolet divergences, provided
the background field has no component along the simple U(1) gauge
group factor. To this end let us use the series of transformations
of section (3) to put the novel contributions in the form
$$\eqalign{ 
{\cal A}^{({\cal R})}(i,j\vert {\cal B})= 
-{i\over 4}   v^{(4)}v^{(2)}
s_{ij} 
(q_i+q_j) {\cal B}& \ 
 \int_0^\infty dl \ \Biggl(
 \sum_{w\in\Gamma_2} e^{-w^2 l /2\pi -i (a_i+a_j)\cdot w } \Biggr)
\times\cr\times &
 {1\over  \eta^{8} }
 { \eta^{2} \over \theta_4^2}  
{\theta_1^\prime (0)
\over \theta_1 ({\epsilon})} 
  \Biggl(
  \theta_3 ({\epsilon})
\theta_3 
\theta_2^2  - 
 \theta_2 ({\epsilon})
\theta_2 
\theta_3^2 
\Biggr) \ \cr}
\eqno(4.9)
$$
$$\eqalign{
{\cal M}^{({\cal R})}(i\vert {\cal B})= 
4i  v^{(4)}v^{(2)}
 q_i {\cal B} &\ 
\int_0^\infty dl\ \Biggl(
 \sum_{w\in\Gamma_2} e^{-2w^2 l /\pi -2i a_i\cdot w } \Biggr)
\times\cr\times &
 {1\over  \eta^{8} }
 { \eta^{2} \over \theta_2^2}  
{\theta_1^\prime (0)
\over \theta_1 ({\epsilon\over 2})} 
  \Biggl(
  \theta_3 ({\epsilon\over 2})
\theta_3 
\theta_4^2  - 
 \theta_4 ({\epsilon\over 2})
\theta_4 
\theta_3^2 
\Biggr) \ \cr}
\eqno(4.10)
$$
and 
$$\eqalign{  
{\cal A}_{ND}(i\vert {\cal B})= 
-{i\over 16 } v^{(4)}v^{(2)}
q_i {\cal B} \ &
 \int_0^\infty dl  \Biggl(
 \sum_{w\in\Gamma_2} e^{-w^2 l /2\pi -i a_i\cdot w } \Biggr)
\times\cr\times &
{1\over  \eta^{8} }
 { \eta^{2} \over \theta_2^2} 
{\theta_1^\prime (0)
\over \theta_1 ({\epsilon})} 
  \Biggl(
  \theta_3 ({\epsilon})
\theta_3 
\theta_4^2  - 
 \theta_4 ({\epsilon})
\theta_4 
\theta_3^2 
\Biggr) \ \cr}
\eqno(4.11)
$$
The elliptic functions in the above expressions are at
modulus $\tau = 2il$ for the annuli and $\tau = 2il -{1\over 2}$
for the M\"obius strip, and it is usefull to recall that
$\epsilon$ is defined by eq. (2.6) with $q_i=q_j$ for the
M\"obius, and $q_j =0$ for the ND contribution. Using 
asymptotic expansions of the
 integrands in the large-$l$ region,
we find the following structure of  divergences
$$
\eqalign{\ \  
{\cal A}^{({\cal R})}(i,j\vert {\cal B})& \sim  
-iv^{(4)}v^{(2)} s_{ij} 
\ 
 \int^\infty dl \times\ 
\Biggl[ 
\underbrace{
 \sqrt{1+q_i^2 \pi^2 {\cal B}^2}
 \sqrt{1+q_j^2 \pi^2 {\cal B}^2}
}_{NS-NS}\ 
+\ 
\underbrace{q_iq_j \pi^2 {\cal B}^2 -1}_{RR}
\Biggr] \ , \cr
\ &\cr
{\cal M}^{({\cal R})}(i\vert {\cal B}) &\sim 
-8i v^{(4)}v^{(2)}
 \int^\infty dl \times\ 
\Biggl[ 
\underbrace{
 { q_i^2\pi^2 {\cal B}^2 \over  
\sqrt{1+q_i^2\pi^2 {\cal B}^2}}
}_{NS-NS}\ 
+\ \  
\underbrace{0}_{RR}
\Biggr] \ , \cr
\ & \cr
{\cal A}_{ND}(i\vert {\cal B})& \sim
{i\over 8} v^{(4)}v^{(2)} 
 \int^\infty dl \times\ 
\Biggl[ 
\underbrace{
 { q_i^2\pi^2 {\cal B}^2 \over  
\sqrt{1+q_i^2\pi^2 {\cal B}^2}}
}_{NS-NS}\ 
+\ \  
\underbrace{0}_{RR}
\Biggr] \cr} 
 \eqno(4.12)
$$
The M\"obius and ND annulus divergences 
 cancel exactly each other,
after summing the latter over  two orientations and  32
possible D-endpoints. Summing over opposite endpoint charges
 in the remaining
annulus diagram, one finds that
 all but a quadratic RR contribution vanish. The final
result therefore reads
$$
\delta{\cal F} \sim -{i} v^{(4)} v^{(2)}
(tr Q)^2
\pi^2 {\cal B}^2 \int^{\infty} dl \ . \eqno(4.13)
$$
with  the trace  in the fundamental
 representation of the gauge group.

This ultraviolet  divergence  comes from  tadpoles of the
twisted RR 4-forms, which couple to the background field through
the generalized Green-Schwarz 
 interaction
 \cite{Mich}
$$ (2\pi)^{-5/2} \sum_I \int d^6x \ {1\over 4!\ 2}
\epsilon^{\mu\nu\rho\sigma\kappa\tau} C^{(I)}_{\mu\nu\rho\sigma}
tr F_{\kappa\tau} \ , \eqno(4.14)
$$
for canonically-normalized 4-forms. 
The coupling gives mass to the $U(1)$ (abelian)  gauge field,
rendering a  background inconsistent. 
For non-abelian background fields, on the other hand, 
 the structure of ultraviolet
divergences is identical to that  of the toroidal
 model:
$${\cal F}^{orbifold} \sim {1\over 2} {\cal F}^{toroidal} \ . $$
 Taking
into account the halving of the volume, we may conclude
in particular  that, 
with SO(32) normalizations for the generators of U(16),
  the relation between gauge and gravitational
couplings stays the same. Furthermore since in the toroidal theory
the gauge coupling is not renormalized, we may conclude that in  the
orbifold the renormalization is ultraviolet finite. 
We will now see explicitly how this comes about.

\vskip 0.9cm

{\bf 5. Gauge-coupling renormalization}
\vskip 0.2cm

In order to extract the quadratic piece in the weak-field
expansion of $\delta{\cal F}$, we will make use of the
identites
$$
\theta_4^{\prime\prime}\theta_4 \theta_3^{\ 2}-
\theta_3^{\prime\prime}\theta_3 \theta_4^{\ 2}
= 4\pi^2 \eta^6 \theta_2^{\ 2} \ , \eqno(5.1)
$$
and 
$$
\theta_3^{\prime\prime}\theta_3 \theta_2^{\ 2}-
\theta_2^{\prime\prime}\theta_2 \theta_3^{\ 2}
= 4\pi^2 \eta^6 \theta_4^{\ 2} \ . \eqno(5.2)
$$
The first of these identities follows from eq. (2.11), if 
one expands to quadratic order around $\nu ={1\over 2}$.
Note  that
a shift of the argument of a $\theta$-function by ${1\over 2}$
can be absorbed into a change of spin structure, and that
 $\theta_1^\prime(0) = 2\pi\eta^3$.
The second identity  is just a modular
 transformation of the first.
Using these two identities, one finds that the 
amplitudes (4.5-7)  expanded to quadratic order in
$\epsilon \simeq (q_i+q_j){\cal B}\ll 1$, collapse to 
contributions  of six-dimensional massless states:
$$\eqalign{
{\cal A}^{(R)} (i,j\vert {\cal B})&= 
-i  V^{(4)} {{\cal B}^2\over 8\pi^2}
  s_{ij} (q_i+q_j)^2
 \int_0^\infty {dt\over t}  
 \Bigl(  \sum_{a_i+a_j+^*\Gamma_2} e^{-{\pi t} p^2/2} \Bigr)
\ +o({\cal B}^4)
\cr
{\cal M}^{(R)} (i\vert {\cal B}) &= 
-i  V^{(4)} {{\cal B}^2\over 2\pi^2}
  q_i^2
 \int_0^\infty {dt\over t}  
 \Bigl(  \sum_{2a_i+^*\Gamma_2} e^{-{\pi t} p^2/2} \Bigr)
\ +o({\cal B}^4) \cr
{\cal A}_{ND} (i\vert {\cal B}) &= 
+i  V^{(4)} {{\cal B}^2\over 32 \pi^2}
  q_i^2
 \int_0^\infty {dt\over t}  
 \Bigl(  \sum_{a_i+^*\Gamma_2} e^{-{\pi t} p^2/2} \Bigr)
\ +o({\cal B}^4)
\cr} \eqno(5.3)
$$
This collapse to the zero-mode space,
noted previously  by Douglas and Li \cite{DL},
 is
to be expected: indeed, as argued in the introduction,
 only BPS states
can  contribute to  threshold effects, and the only
BPS states of the  open string are the massless
(before Higgsing)  modes of the
 six-dimensional model. 
A similar result is well known  in 
 the heterotic string \cite{greg}
where, however, the spectrum of
  BPS states includes infinite string excitations with no
simple field-theoretic description.

Putting together eqs. (4.4) and (5.3) we arrive at the following
expression for the full free energy, including classical and
one-loop contributions

$$
\eqalign{
\  {\cal F}({\cal B})/& V^{(4)} =\  {{\cal B}^2 \over 2 g^2_{(4)}}  
 +\ {{\cal B}^2\over 8\pi^2}\   
 \int_0^\infty {dt\over t}\ \times
\Biggl[ \sum_i  q_i^2 \Bigl(\sum_{a_i+^*\Gamma_2}
 4  e^{-{\pi t} p^2/2}- \cr &
-  \sum_{2a_i+^*\Gamma_2}
  e^{-{\pi t} p^2/2}\Bigr)
- \sum_{ij} 
  s_{ij} (q_i+q_j)^2\sum_{a_i+a_j+^*\Gamma_2}
  {1\over 4}  e^{-{\pi t} p^2/2}\ \Biggr]
 +o({\cal B}^4)  \cr}  \eqno(5.4)
$$

\noindent where the SU(16) generators are 
 normalized to $tr_{16} Q^2 = {1\over 2}$,
and we recall that the Chan-Patton charges run over both the 
$16$ and the ${\overline 16}$ representations separately.  
As a check let us extract the leading infrared divergence of
the coupling-constant renormalization
  in the limit of vanishing Wilson lines. Cutting
off $t < 1/\mu^2$ one finds after some straightforward algebra
$$
{4\pi^2\over g_{(4)}^2} \Biggl\vert_{1-loop} =
{4\pi^2\over g_{(4)}^2} \Biggr\vert_{tree} - 6\ 
 log\mu +\  {\rm IR\ finite}
\ , \eqno(5.5)
$$
in agreement with the correct $\beta$-function coefficient of the
N=2 theory in four dimensions,
$$
C_{adj} - 2 C_{120} - 16 C_{fund} = -6 \ . 
$$
Put differently, expression (5.4) correctly reproduces
 the logarithmic
part of the one-loop prepotential for this model.
\vskip 0.2cm

This expression  is formally identical to  that
of  Kaluza-Klein theory  compactified from
six to four dimensions.   
If we were to  impose a uniform  ultraviolet
 momentum cutoff,  $ t> 1/ \Lambda^{2}$, the result would
therefore  be 
 quadratically
divergent. The cutoff dictated by string theory is,  however,  much
smarter! It is uniform in  transverse time $l$, 
which means that if we cutoff the  annulus at $ t= 1/ \Lambda^{2}$,
we must cutoff   the M\"obius strip 
 at  $t= 1/ 4\Lambda^2$. 
To render finiteness more explicit, let us
perform the Poisson
 resummations  to the transverse channel, and put  eq. (5.4)
in the form
$$
\eqalign{
\delta{\cal F}({\cal B}) =
2\pi^2 {\cal B}^2 v^{(4)} & v^{(2)} \   
 \int_0^\infty dl\ \sum_{w\in \Gamma_2}
\Biggl[ \sum_i   q_i^2 \Bigl(
   e^{-w^2l/2\pi+iw\cdot a_i} 
-  e^{-2w^2l/\pi+2iw\cdot a_i}\Bigr)\cr
-& {1\over 16}  \sum_{ij} 
  s_{ij} (q_i+q_j)^2  e^{-w^2l/2\pi+iw\cdot( a_i+a_j)}
\ \Biggr]\ + o({\cal B}^4) \cr}  \eqno(5.6)
$$
We may now  perform  the $l$-integrations,
 and sum over the
two complex conjugate representations $16$ and ${\overline  {16}}$
over which $i,j$ run.
 Since generic Wilson-line backgrounds  break the gauge symmetry
to the Cartan subgroup of SU(16), the answer
  is better expressed as
the one-loop correction, $\Delta_{ij}$, 
 to the gauge  kinetic function
defined through
$$
{\cal L}_{eff} = 
({1\over 2g_{(4)}^2}\delta_{ij} + \Delta_{ij})
 F^{i, \mu\nu} F^j_{ \mu\nu} \eqno(5.7)
$$
where $F_{\mu\nu}^i$ is a traceless diagonal 16-dimensional 
 matrix.
 The final result reads
$$\eqalign{
\Delta_{ij}&(a,\Gamma_2) = 
 \sum_{w\in\Gamma_2-\{0\}}\  {v^{(2)}\over 2\pi w^2}\ 
\Biggl[\  cos(w\cdot a_i) cos(w\cdot a_j) + \cr &+
\delta_{ij}\  \Bigl\{  4 cos(w\cdot a_i) - cos(2w\cdot a_i)
-sin(w\cdot a_i) \sum_k sin(w\cdot a_k) \Bigr\}  \Biggr]
\cr}  \eqno(5.8)
$$
where in a generic point of moduli space
 the winding sums are  manifestly convergent.
Notice that in addition to the Wilson lines,
 the above result gives  the  dependence
on  the moduli $ ImT$ and $ U$  of the
2-torus. It should be furthermore straightfoward to extend the
analysis so as to account for the  six-dimensional gauge  moduli.
 The dependence on $ReT$, 
which in open-string theory  is a RR 2-form background,
may however be harder  to extract.

\vskip 0.9cm
{\bf 6. Concluding Remarks}
\vskip 0.2cm

Our  calculation
 illustrates in a very simple context the way in which string
theory produces  finite answers: in the case at hand it
is simply field theory but with a very smart  cutoff on the
momenta. These results have non-trivial implications, both
in the context of heterotic-type-I duality \cite{het-I},
and for the study of moduli spaces of D-branes \cite{dyn,DL}.
In this latter context in particular it implies  
that  the metric in the moduli space 
of N=2 configurations of D-branes
 is given  entirely 
 by simple  and  massless 
 closed-string exchange.
We plan to pursue these issues further in the near future.

\vskip 1cm
\noindent
{\bf Acknowledgments} \\
We thank E. Kiritsis, C. Kounnas, N. Obers,
  H. Partouche, G. Pradisi,
A. Sagnotti
and P. Vanhove  for usefull
conversations during various stages of the work,
 which  was supported
by EEC grant CHRX-CT93-0340.

\vfil\eject

\vskip 0.9 cm
{\bf Appendix}
\vskip 0.2cm

The field-independent contributions of closed and DD strings to
the one-loop free energy are given by the following amplitudes
$$\eqalign{
i {\cal F}_{closed} = {1\over 4} ( {\cal T}&+{\cal T}^{({\cal R})}
+{\cal K} +  {\cal K}^{({\cal R})})+\cr
&+{1\over 4} ( 
{\cal T}_{twist} +  {\cal T}_{twist}^{({\cal R})}+
{\cal K}_{twist}+{\cal K}_{twist}^{({\cal R})}
)\cr}\eqno(A.1) 
$$
and 
$$
i {\cal F}_{DD} =  {1\over 4} ( {\cal A}_{DD} +
 {\cal A}_{DD}^{({\cal R})}
+ {\cal  M}_D+  {\cal M}^{({\cal R})}_{D})
\ , \eqno(A.2)
$$
where summation over Dirichlet  endpoint states is implicitly
performed in the second line.
The torus amplitudes are well known from the type-II string, and
are by themselves ultraviolet finite. ${\cal K}$ was computed
in section 2, while the remaining Klein bottle amplitudes read
in six uncompactified dimensions
$$
{\cal K}^{({\cal R})}=
-{i\over 2} 
 V^{(6)} \int_0^\infty {dt\over t} (2\pi^2 t)^{-3}
 \Bigl( \sum_{w\in \Gamma_4} e^{-{ t} w^2/2\pi } \Bigr)
{1\over  \eta^{12} } \times  
\sum_{\alpha=2,3,4}  \  {1\over 2} s_\alpha
\theta_\alpha^4   \eqno(A.3)
$$
and 
$$
{\cal K}_{twist} +{\cal K}_{twist}^{({\cal R})} =
-{i \over 2}  16 
 V^{(6)} \int_0^\infty {dt\over t} (2\pi^2 t)^{-3}
{1\over  \eta^{8}}{\eta^2\over \theta_4^2}
(\theta_3^2 \theta_2^2 - \theta_2^2\theta_3^2) \eqno(A.4)
$$
The
argument of all modular functions is $\tau = 2it$, while
the factor 16 appearing in the second line counts
the number of  fixed-points of the orbifold.
Notice that the  role of 
${\cal K}^{({\cal R})}$ is to symmetrize
  winding Neveu-Schwarz, and antisymmetrize winding RR
states, since pure winding is left unchanged by the combined
action of the operator $\Omega{\cal R}$. The role of
${\cal K}_{twist}+{\cal K}_{twist}^{({\cal R})}$ on the
other hand is to symmetrize twisted NS-NS
 states, and antisymmetrize twisted RR states,
 giving a net number
of  respectively 48 and 16
massless  space-time singlets \cite{GP}.
The DD amplitudes, assuming that
 all 32 5-branes are at the same fixed point,   read
$$
{\cal A}_{DD} = 
- 2^{10}\times {i\over 2} 
 V^{(6)} \int_0^\infty {dt\over t} (2\pi^2 t)^{-3}
 \Bigl( \sum_{w\in \Gamma_4} e^{-{ t} w^2/2\pi } \Bigr)
{1\over  \eta^{12} } \times  
\sum_{\alpha=2,3,4}  \  {1\over 2} s_\alpha
\theta_\alpha^4   \eqno(A.5)
$$
and
$$
{\cal M}_{D}^{({\cal R})} =
 2^{5}\times {i\over 2} 
 V^{(6)} \int_0^\infty {dt\over t} (2\pi^2 t)^{-3}
 \Bigl( \sum_{w\in \Gamma_4} e^{-{ t} w^2/2\pi } \Bigr)
{1\over  \eta^{12} } \times  
\sum_{\alpha=2,3,4}  \  {1\over 2} s_\alpha
\theta_\alpha^4   \eqno(A.6)
$$
where  $\tau = {it\over 2}$  for the annulus and
$\tau ={it\over 2}+{1\over 2}$ for the M\"obius strip, as usual.
The other two contributions are zero in the Neveu-Schwarz and
Ramond sectors separately. For
 ${\cal A}_{DD}^{({\cal R})}$ this is obvious since it
 is proportional to $(tr \gamma_{5,{\cal R}})^2$.
As for
 ${\cal M}_{D}$, the reason is  more subtle: 
it can be traced to the fact that the action of $\Omega$ on a
DD (super)coordinate has an extra minus sign compared to its action
on a NN (super)coordinate \cite{GP}. 
As a result  $\Omega$ anticommutes with
 the fermionic DD zero
modes in the Ramond sector, so that the
 corresponding contribution
in the M\"obius amplitude vanishes. Space-time supersymmetry then
ensures that the contribution of Neveu-Schwarz states is also zero.

When transforming to the cross 
 $l$-channel, ${\cal K}^{({\cal R})}$,
${\cal A}_{DD}$ and  ${\cal M}_{D}^{({\cal R})}$ give unphysical
tadpoles proportional to the inverse volume of the orbifold.
These cancel between the three diagrams \cite{GP}, a phenomenon
that is T-dual to the usual cancellation between the  NN annulus
and  M\"obius
strip,  and the (untwisted) Klein bottle.
  Notice that this T-duality
exchanges the orientation-reversing operator  $\Omega$ with 
 $\Omega{\cal R}$. Finally  the twisted Klein
 bottles
vanish in Neveu-Schwarz or
RR $l$-channels separately, and hence do not
 create any anomalies.
Conversely, this shows that in the direct channel, we are
forced to antisymmetrize RR twisted states,
 if we have symmetrized
the NS-NS ones.

\vfil\eject

\end{document}